\documentclass[12pt]{article}
\usepackage{amsmath}
\usepackage{amssymb}
\usepackage{graphicx}
\usepackage{epstopdf}
\usepackage{cite}
\usepackage{hyperref}
\newcommand{\beginsupplement}{\setcounter{equation}{0}
\renewcommand{\theequation}{S\arabic{equation}}\setcounter{figure}{0}
\renewcommand{\thefigure}{S\arabic{figure}}  }
\begin{document}
\title{Cross-linking patterns and their images in swollen and deformed gels}

\author{Sergey Panyukov\\{\it P. N. Lebedev Physics Institute, Russian Academy of Sciences,}\\{\it Moscow 117924, Russia}
\and  Yitzhak Rabin\\ {\it Department of Physics and Institute of
Nanotechnology} \\ {\it and Advanced Materials,} \\{\it Bar-Ilan University, Ramat-Gan 52900,
Israel}}
\maketitle

\begin{abstract}
Using the theory of elasticity of polymer gels we show that large-scale
\textit{cross-link} density patterns written into the structure of the network
in the melt state, can be revealed upon swelling by monitoring the
\textit{monomer} density patterns. We find that while isotropic deformations
in good solvent yield magnified images of the original pattern, anisotropic
deformations distort the image (both types of deformation yield affinely
stretched images in $\theta$ solvents). We show that in ordinary solids with
spatially inhomogeneous profile of the shear modulus, isotropic stretching
leads to distorted density image of this profile under isotropic deformation.
Using simple physical arguments we demonstrate that the different response to
isotropic stretching stems from fundamental differences between the theory of
elasticity of solids and that of gels. Possible tests of our predictions and
some potential applications are discussed.
\end{abstract}
\section{Introduction}

Polymer networks are unique soft solids which can be significantly deformed
without irreversible damage to their structure. A network is formed by
cross-linking a melt or a semidilute solution of polymer chains. Once a
homogeneous (on length scales large compared to its ``mesh'' size) network is
formed, one can generate large-scale patterns in it by further cross-linking,
followed by swelling (and possibly stretching) of the network, resulting in a
gel inhomogeously swollen by solvent. This can be done, for example, by adding
light-sensitive cross-links to a transparent network. Focusing a laser beam in
regions inside the gel one can \textquotedblleft write\textquotedblright%
\ information into gel structure in the form of $2D$ or $3D$ patterns of
cross-linking density. In this paper we show that although such information is
hidden at preparation conditions, it can be recovered by swelling the gel
since unobservable variations of cross-link density in the melt are
transformed into observable variations of monomer density in the swollen gel.

Regions of a gel with increased cross-link concentration can be considered as
inclusions with enhanced elastic modulus. If such inclusions deform
differently from polymer matrix, as in case of any normal elastic solids, they
would induce elastic stresses in the gel and initial pattern would be
significantly distorted due to long range character of elastic interactions.
This scenario determines, for example, the elastic properties of amorphous
polycrystalline solids but it does not apply to polymer gels, because of the
unusual character of gel elasticity. We show that in swollen gels that are
isotropically stretched by absorption of solvent, the observed \textit{monomer
density pattern} is not distorted and is simply an affinely stretched variant
of the \textit{initial cross-linking pattern}. Such gels can serve as a
magnifying glass that enlarges the initially written pattern without distorting
its shape. The corresponding magnification factor can be very large in case of
super-elasic networks.

\section{Free energy of a gel with cross-linking density pattern}

\begin{figure}[tbh]%
\centering
\includegraphics[
height=1.8743in,
width=3.2812in
]%
{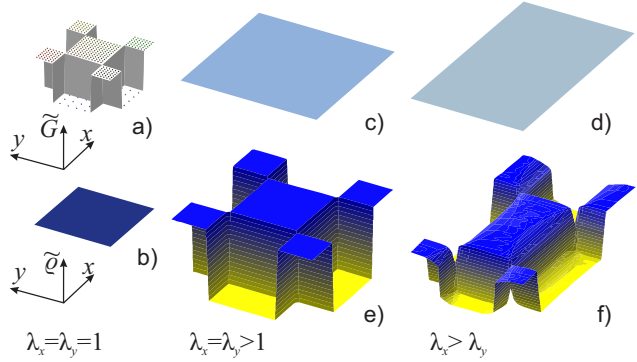}%
\caption{Initial cross-link concentration $c\left(  x,y\right)  $ a); monomer
density profiles $\rho\left(  x,y\right)  $: of a gel in a melt state b), in
reference state after stretching c), d), in stretched equilibrium state e),f).
Gel is isotropically stretched by factors $\lambda_{1}=\lambda_{2}=1.5$ c), e)
and anysotropically stretched by factors $\lambda_{1}=2,\lambda_{2}=1$ d), f).
}%
\label{Pattern}%
\end{figure}

In this paper we use the simplest mean field model of a gel with free energy
 \cite{Onuki,inhomogeneous}:%
\begin{equation}
A=\int\left(  \frac{G\left(  \mathbf{x}_{0}\right)  }{2}\sum_{ij}F_{ij}%
^{2}+f\left[  \rho\left(  \mathbf{x}_{0}\right)  \right]  \right)
d\mathbf{x}_{0} \label{Ai} 
\end{equation}
Here $\rho\left(  \mathbf{x}_{0}\right)  $ is monomer density as function of
coordinates $\mathbf{x}_{0}$ in preparation state. We assume that the gel was
initially cross-linked in a polymer melt and then a pre-programmed pattern in
cross-link concentration (i.e., a well-defined region of higher cross-link
density compared to that of the surrounding network) is created in the network
using, say, a light-sensitive cross-linking technique (the case of
cross-linking in semi-dilute solution in good solvent is analyzed in SI). Here
$G\left(  \mathbf{x}_{0}\right)  $ is the polymer contribution to the elastic
modulus of the cross-linked melt (which is proportional to the local
cross-link density)%
\begin{equation}
G\left(  \mathbf{x}_{0}\right)  =\bar{G}+\tilde{G}\left(  \mathbf{x}%
_{0}\right)
\end{equation}
and $\tilde{G}\left(  \mathbf{x}_{0}\right)  $ represents the variations of
cross-link density introduced by the second cross-linking step
(Fig.~\ref{Pattern}a).
$f\left(  \rho\right)  $ is the osmotic (interaction) part of the free energy
of the gel, with monomer density $\rho$. $F$ is the deformation gradient
tensor%
\begin{equation}
F_{ij}=\frac{\partial x_{i}}{\partial x_{0j}} \label{Fij}%
\end{equation}
and $\mathbf{x}$ are coordinates of deformed gel. It is convenient to assume
that the gel is deformed with respect to preparation state in two stages:%
\begin{equation}
F_{ij}=\sum_{l}F_{il}^{u}F_{lj}^{\lambda}%
\end{equation}
Thus, the gel is first stretched with respect to preparation state by factors
$\lambda_{i}$ along axes $i$. For such a deformation $\mathbf{x}=\lambda
\cdot\mathbf{x}_{0}$ with components $x_{i}=\lambda_{i}x_{0i}$ and we get
\begin{equation}
F_{ij}^{\lambda}=\lambda_{i}\delta_{ij},\qquad\rho=\bar{\rho}=\frac{\rho_{0}%
}{\prod_{i}\lambda_{i}} \label{Fl}%
\end{equation}
where $\rho_{0}$ is the uniform monomer density in the undeformed state of
preparation (Fig.~\ref{Pattern}b). Notice that the coordinates $\mathbf{x}$
describe a stretched network with inhomogeneous cross-link density but a
\textit{{homogeneous monomer density}} (Figs.~\ref{Pattern}c and d).

Even though such a homogeneous (in monomer density) state does not minimize
the free energy and therefore is not an equilibrium state of the deformed gel,
we use it as a reference state. The true equilibrium state of the deformed
network has an inhomogeneous monomer density profile and is defined by
introducing a displacement field $\mathbf{u}\left(  \mathbf{x}\right)  $
defined with respect to the above reference state:
\begin{equation}
\mathbf{x}^{\prime}=\mathbf{x}+\mathbf{u}\left(  \mathbf{x}\right)
\end{equation}
and we get gradient tensor and monomer density as function of coordinates
$\mathbf{x}$%
\begin{align}
F_{ij}^{u} &  =\delta_{ij}+\frac{\partial u_{i}}{\partial x_{j}},\qquad
\rho\left(  \mathbf{x}\right)  =\frac{\bar{\rho}}{\det\left(  F_{ij}%
^{u}\right)  },\label{F}\\
\det\left(  F_{ij}^{u}\right)   &  \simeq1+\sum_{i}\frac{\partial u_{i}%
}{\partial x_{i}}\label{detF}%
\end{align}

Minimizing the free energy in Eq.~(\ref{Ai}) with respect to displacements
$\mathbf{u}$ at the preparation state (all $\lambda_{i}=1$) we conclude that
in a melt the cross-links and the monomers will remain at their previous
possitions and the elastic reference state will not change after relaxation.
We conclude that information about the pattern written on network structure is
hidden in preparation state and can only be revealed after swelling.

\section{What is the equilibrium density profile?\label{GOOD}}

In a swollen state the monomer density is small and the interaction energy can
be expanded as $f\left(  \rho\right)  \simeq k_{B}TB\rho^{2}/2$, where $k_{B}$
is Boltzmann constant, $T$ is temperature and $B$ is second virial
coefficient. Expanding the free energy in powers of $\mathbf{u}$ and
integrating over the volume of the undeformed network with measure
$d\mathbf{x}_{0}=d\mathbf{x}/\prod_{i}\lambda_{i}$ we get%
\begin{align}
\Delta A  &  =\int\left[  \tilde{G}\left(  \lambda^{-1}\mathbf{\cdot
x}\right)  \sum_{i}\lambda_{i}^{2}\frac{\partial u_{i}}{\partial x_{i}}%
+\frac{\bar{G}}{2}\sum_{ij}\left(  \lambda_{j}\frac{\partial u_{i}}{\partial
x_{j}}\right)  ^{2}\right. \nonumber\\
&  \left.  +\frac{K_{\text{os}}}{2}\left(  \sum_{i}\frac{\partial u_{i}%
}{\partial x_{i}}\right)  ^{2}\right]  \frac{d\mathbf{x}}{\prod_{i}\lambda
_{i}},\quad K_{\text{os}}=k_{B}TB\bar{\rho}^{2}. \label{A}%
\end{align}

The equilibrium deformation of the gel is found by minimizing this free
energy. Its variation is%
\begin{align}
\delta A  &  =\int\left(  -\sum_{i}\delta u_{i}\lambda_{i}^{2}\frac
{\partial\tilde{G}}{\partial x_{i}}-\bar{G}\sum_{ij}\lambda_{j}^{2}%
\frac{\partial^{2}u_{i}}{\partial x_{j}^{2}}\delta u_{i}\right. \nonumber\\
&  \left.  -K_{\text{os}}\sum_{ij}\frac{\partial^{2}u_{j}}{\partial
x_{i}\partial x_{j}}\delta u_{i}\right)  \frac{d\mathbf{x}}{\prod_{i}%
\lambda_{i}}%
\end{align}
and therefore, the minimum condition is%
\begin{equation}
-\lambda_{i}^{2}\frac{\partial\tilde{G}}{\partial x_{i}}-\bar{G}\sum
_{j}\lambda_{j}^{2}\frac{\partial^{2}u_{i}}{\partial x_{j}^{2}}-K_{\text{os}%
}\sum_{j}\frac{\partial^{2}u_{j}}{\partial x_{i}\partial x_{j}}=0 \label{Gu}%
\end{equation}
We are interested only in variations of monomer density%
\begin{equation}
\frac{\tilde{\rho}\left(  \mathbf{x}\right)  }{\bar{\rho}}\simeq-\sum_{i}%
\frac{\partial u_{i}\left(  \mathbf{x}\right)  }{\partial x_{i}}%
\end{equation}
where $\bar{\rho}$ is average density. Taking the gradient of both sides of
Eq.~(\ref{Gu}) we obtain an equation for the variations of monomer density%
\begin{equation}
\sum_{i}\lambda_{i}^{2}\frac{\partial^{2}\tilde{G}\left(  \lambda
^{-1}\mathbf{\cdot x}\right)  }{\partial x_{i}^{2}}-\sum_{i}\gamma_{i}%
^{2}\frac{\partial^{2}}{\partial x_{i}^{2}}\frac{\tilde{\rho}\left(
\mathbf{x}\right)  }{\bar{\rho}}=0 \label{d2}%
\end{equation}
where%
\begin{equation}
\gamma_{i}^{2}=\bar{G}\lambda_{i}^{2}+K_{\text{os}} \label{gamma}%
\end{equation}
The solution of this equation%
\begin{equation}
\frac{\tilde{\rho}\left(  \mathbf{x}\right)  }{\bar{\rho}}=\sum_{i}\lambda
_{i}^{2}\frac{\partial^{2}}{\partial x_{i}^{2}}\int g\left[  \gamma^{-1}%
\cdot\left(  \mathbf{x}-\mathbf{y}\right)  \right]  \tilde{G}\left(
\lambda^{-1}\mathbf{\cdot x}\right)  \frac{d\mathbf{y}}{\prod_{j}\gamma_{j}}
\label{rG}%
\end{equation}
is expressed through Green's functions of the Laplace equation in 2 and 3
dimensions, respectively:%
\begin{equation}
g_{2D}\left(  \mathbf{x}\right)  =\frac{1}{4\pi}\ln\sum_{i}x_{i}^{2},\qquad
g_{3D}\left(  \mathbf{x}\right)  =-\frac{1}{4\pi\sqrt{\sum_{i}x_{i}^{2}}}
\label{g}%
\end{equation}

In case of isotropically stretched/swollen gel with all $\lambda_{i}=\lambda$
the equilibrium monomer density depends on local cross-link concentration,%
\begin{equation}
\frac{\tilde{\rho}\left(  \mathbf{x}\right)  }{\bar{\rho}}=\frac{\lambda^{2}%
}{\gamma^{2}}\tilde{G}\left(  \frac{\mathbf{x}}{\lambda}\right)  \label{dr}%
\end{equation}
We conclude that under isotropic deformation such as swelling, the monomer
density produces an undistorted, uniformly stretched image of the pattern of
cross-link density originally \textquotedblleft written\textquotedblright\ on
the homogeneous network (compare Figs. \ref{Pattern}a and e).

Equilibrium displacement is expressed through the variation of monomer
density, Eq.~(\ref{dr}), as%
\begin{equation}
u_{i}\left(  \mathbf{x}\right)  =-\frac{\partial}{\partial x_{i}}\int g\left(
\mathbf{\mathbf{x}-y}\right)  \frac{\tilde{\rho}\left(  \mathbf{y}\right)
}{\bar{\rho}}d\mathbf{y} \label{dux}%
\end{equation}
We conclude that although density variations in isotropically deformed gels
are strictly local, there is long-range strain field decaying as power law of
the distance $\left\vert \mathbf{\mathbf{x}-y}\right\vert $. This strain
induces a stress distribution in the gel, which can be observed by measuring
the birefringence of transmitted light (stress-optical law
\cite{stress-optical}).

In anisotropically deformed networks the pattern is strongly distorted
(compare Figs. \ref{Pattern}a and f) and $\tilde{\rho}\left(  \mathbf{x}%
\right)  $ decays as power law of a distance $\left\vert \mathbf{x}%
-\mathbf{y}\right\vert $ from the localized cross-link density inhomogeneity
$\tilde{G}\left(  \mathbf{y}\right)  $. Observe that variations of monomer
density are largest along the direction of stretching. This effect is closely
related to the well known \textquotedblleft butterfly\textquotedblright%
\ picture in contour plots of neutron scattering from random inhomogeneities
of network structure in anisotropely deformed swollen gels \cite{bastide}.

\section{What is the difference between deformation of gels and elastic
solids?}

In order to understand the difference between gels and normal solids we recall
that the free energy of any solid is a functional of the nonlinear strain
tensor $u_{ik}$ \cite{Landau},%
\begin{equation}
2u_{ik}=\sum_{j}F_{ij}F_{kj}-\delta_{ik}=\frac{\partial u_{i}}{\partial x_{k}%
}+\frac{\partial u_{k}}{\partial x_{i}}+\sum_{j}\frac{\partial u_{i}}{\partial
x_{j}}\frac{\partial u_{k}}{\partial x_{j}} \label{uik}%
\end{equation}
While the last term is usually neglected in the linear theory elasticity of
solids because solids behave elastically only under small deformations, it can
be shown that only this nonlinear term contributes to the elasticity of gels
and that the elastic part of the free energy of gels (Eq. (\ref{Ai})) is
linear in this nonlinear strain\cite{PhysRep,Alexander}. Physically, the
difference between elastic energy \textit{of a solid,} which is a quadratic
form in the linear strain and\textit{ of a gel,} which is linear in the
nonlinear strain tensor, stems from the fact that while in solids there is a
stress-free state of equilibrium (crystal lattice) that minimizes the energy
of interaction between the atoms, the equilibrium state of gels is not
stress-free. Polymer networks are made of entropic springs and, in the absence
of osmotic pressure due to permeation by good solvent or due to excluded
volume interactions in the melt state, such networks would collapse to the
size of a single spring. The finite length of entropic springs in the swollen
gel is the result of osmotic pressure which can be replaced by equivalent
isotropic stretching forces that act on the outer boundaries of the gel
\cite{Flory-Rehner}.

The difference between gels and solids becomes apparent when considering two
simple toy models of heterogeneous gel and solid as two Hookean springs with
moduli $k_{1}$ and $k_{2}$, connected in series as in Fig.~\ref{Toy}:
\begin{figure}[tbh]%
\centering
\includegraphics[
height=1.5509in,
width=3.2812in
]%
{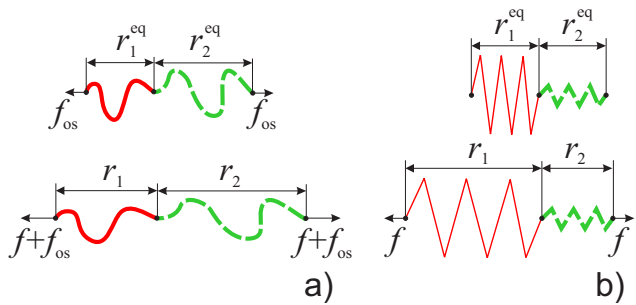}%
\caption{Two springs models, demonstrating affine deformation of gels a) and
non-affine deformation of solids b).}%
\label{Toy}%
\end{figure}

\textit{a) Gel model:} Osmotic pressure is represented by a force
$f_{\text{os}}$ applied to free ends of the connected springs. In the presence
of this force the equilibrium lengths of the Gaussian springs become
$r_{1}^{\text{eq}}=f_{\text{os}}/k_{1}$ and $r_{2}^{\text{eq}}=f_{\text{os}%
}/k_{2}$, and if we apply additional force $f$, each of the springs will
deform affinally with distance $r_{1}+r_{2}$ between the ends to which the
force is applied \noindent(\textquotedblright boundaries\textquotedblright\ of
the system):
\begin{equation}
r_{1}=\lambda r_{1}^{\text{eq}},\quad r_{2}=\lambda r_{2}^{\text{eq}}%
,\quad\lambda=\frac{r_{1}+r_{2}}{r_{1}^{\text{eq}}+r_{1}^{\text{eq}}}%
\end{equation}

\textit{b) Solid model:} The springs of a \textquotedblleft
solid\textquotedblright\ have equilibrium lengths $r_{1}^{\text{eq}}$ and
$r_{2}^{\text{eq}}$ in the stress-free state. During stretching due to force
$f$ applied to the ends of the two-spring system, such a solid deforms
nonaffinelly:%
\begin{equation}%
\begin{array}
[c]{c}%
r_{1}=r_{1}^{\text{eq}}+\left(  \lambda-1\right)  \left(  r_{1}^{\text{eq}%
}+r_{2}^{\text{eq}}\right)  \dfrac{k_{2}}{k_{1}+k_{2}},\\
r_{2}=r_{2}^{\text{eq}}+\left(  \lambda-1\right)  \left(  r_{1}^{\text{eq}%
}+r_{2}^{\text{eq}}\right)  \dfrac{k_{1}}{k_{1}+k_{2}}%
\end{array}
\end{equation}
with the soft spring ($k_{1}<k_{2}$) stretched more than the rigid one.

These two simple toy models illustrate why under isotropic deformations,
cross-linking density patterns in gels are stretched affinally, whereas soft
regions in solids would undergo larger deformation compared to more rigid
regions, thus distorting the original pattern.

\section{Discussion\label{DISCUS}}

We studied the combined effect of swelling and deformation on inhomogeneous
networks, prepared by cross-linking a melt of polymer chains. It is well-known
that cross-link density heterogeneities that have no effect on the monomer
density in the state of preparation (a melt or a concentrated polymer
solution), can be revealed by swelling the gel and observing the enhancement
of light, x-ray and neutron scattering from the resulting monomer density
inhomogeneities\cite{Candau,Geisler,Shibayama,BastideCandau}. In this paper we
focused on a related phenomenon, namely that when large-scale cross-link
density patterns are written into the network structure, the hidden image can
be revealed by swelling and stretching the gel and observing the corresponding
patterns of monomer density. Using the mean field theory of elasticity of
polymer gels we showed that stretching/swelling in good solvent acts as a
magnifying glass: while isotropic stretching reproduces an enlarged but
otherwise undistorted version of the original pattern, anisotropic stretching
distorts this pattern, see figure~\ref{Pattern}.

We compared these results with those obtained for ordinary elastic solids with
inhomogeneous elastic moduli and found that in this case even isotropic
deformations lead to distorted patterns. We showed that the fundamental
difference between response of inhomogeneous gels and solids to isotropic
stretching can be traced back to the fact that unlike regular springs that
have an equilibrium length even in the absence of stress, the equilibrium
length of entropic springs is entirely determined by the osmotic forces that
isotropically stretch the polymer gel.

Finally, we would like to comment on the possibility of experimental
verification and on possible applications of our results. In most application
involving gels such as biomimetic sensors, actuators and artificial muscles
 \cite{Smith}, macroscopically inhomogeneous (layered) gels undergo
shape transitions when the thermodynamic conditions are changed or in response
to application of external fields \cite{Sharon}. In our case, the cross-link
density pattern imprinted into the gel structure by, say, activation of
light-sensitive cross-links, can be microscopic (micron size) and therefore
would have little effect on the shape of the gel. Upon swelling and/or
isotropic stretching in good solvent, the magnified density pattern can be
imaged on a light-sensitive screen. The contrast can be significantly enhanced
by stretching the entire gel in poor solvent or by focusing a laser beam on
the localized pattern and heating it, resulting in local change of the quality
of solvent. Finally, the sensitivity of the image to quality of solvent (the
distortion under anisotropic deformation disappears in $\theta$ solvent - see
SI) can be useful for sensor devices. \vspace{3mm}

\noindent\textbf{Acknowledgments}

\noindent YR's research was supported by the I-CORE Program of the Planning
and Budgeting committee and the Israel Science Foundation, and by the
US-Israel Binational Science Foundation. \vspace{3mm}

\noindent\textbf{Supporting Information.} 

\noindent In the SI we show that patterns obtained by cross-linking a semi-dilute
polymer solution, deform affinely (non-affinely) under isotropic (anisotropic)
deformation, just like in the case of cross-linking in the melt. We then
analyze how the pattern deforms under several different solvent conditions. We
show that the pattern always stretches affinely in a $\theta$-solvent, even
under anisotropic deformations. Since the contrast between the high and the
low monomer density regions can be significantly enhanced in a poor solvent we
proceed to analyze the density profiles in gels that are isotropically
stretched in mildly poor solvents (at lower solubility, such stretched gels
will undergo a transition into a strongly inhomogeneous state characterized by
the appearance of dense filamentous structures \cite{orit}). We find that when
the amplitude of cross-link density variations is sufficiently low, the image
stretches affinely with the isotropic deformation but that for larger density
contrasts the image becomes distorted, especially near the edges and corners
of the pattern.

\newpage

\beginsupplement

\bigskip\textbf{Supplementary Information to:}

\bigskip{\Large \textbf{Cross-linking patterns and their images in swollen and
deformed gels}}

\section*{SI 1. Gel free energy in a good solvent}

Consider a gel prepared in a good solvent at the monomer density $\rho_{0}$
that is swollen to density $\rho$. Its free energy is the sum of elastic and
osmotic contributions. The osmotic pressure $\pi$ of the gel in a good solvent
increases proportionally to the $9/4$ power of monomer density $\rho$
\cite{SI_GeGennes}%
\begin{equation}
\pi\simeq\left(  k_{B}T/b^{3}\right)  {\left(  \rho b^{3}\right)  }%
^{9/4}\label{SI_pi}%
\end{equation}
where $k_{B}$ is Boltzmann constant, $T$ is temperature and $b$ is monomer
size. The osmotic part $A_{\text{os}}$ of the free energy per polymer chain
between network junctions is proportional to the free energy density ($\sim
\pi$) divided by the number of chains per unit volume ($\rho/N$), where $N$ is
the chain degree of polymerization:%
\begin{equation}
A_{\text{os}}^{\text{ch}}\simeq\pi\left/  \left(  \rho/N\right)  \right.
\simeq k_{B}TN{\left(  \rho b^{3}\right)  }^{5/4}\label{SI_osm}%
\end{equation}

The dimension of the chain along the main axis $\alpha$ of deformation is
$R_{\alpha}={\Lambda}_{\alpha}{R}_{0}$,\textit{\ }where ${\Lambda}_{\alpha}$
is deformation factor along this axis (defined as eigenvalue of local
deformation gradient tensor $F$) and ${R}_{0}$ is the chain size in the state
at which the gel was formed. The elastic free energy per chain is%
\begin{equation}
A_{\text{el}}^{\text{ch}}\simeq k_{B}T\sum_{\alpha}{{\left(  \frac{{R}%
_{\alpha}}{{R}_{fl}}\right)  }^{2}}\simeq\ k_{B}T{{\left(  \frac{{R}_{0}}%
{{R}_{fl}}\right)  }^{2}}\sum_{\alpha}{\Lambda}_{\alpha}^{2}\label{SI_el}%
\end{equation}
where ${R}_{fl}$ is the amplitude of fluctuations of the chain in the deformed
state. In a heterogeneous network the direction of the triad of deformation
axes $\alpha$ depends on its position, and the sum of squares of local
deformation factors in Eq.~(\ref{SI_el}) can be rewritten through the deformation
gradient tensor $F$ as
\begin{equation}
\sum_{\alpha}{\Lambda}_{\alpha}^{2}=\sum_{ij}F_{ij}^{2}.\label{SI_l2}%
\end{equation}
Since the mean-square amplitude of chain fluctuations is proportional to the
mean-square polymer size at semi-dilute good solvent conditions and scales
with monomer density as\cite{SI_GeGennes} ${R}_{fl}^{2}\sim\rho^{-1/4}$, while
the mean-square chain size in the preparation conditions scales as ${R}%
_{0}^{2}\sim\rho_{0}^{-1/4}$, the elastic free energy per chain is%
\begin{equation}
A_{\text{el}}^{\text{ch}}\ \simeq\ k_{B}T{\left(  \frac{\rho}{\rho_{0}%
}\right)  }^{1/4}\ \sum_{ij}F_{ij}^{2}\label{SI_el1}%
\end{equation}

At the equilibrium swelling ($\rho=\rho_{\text{eq}}$) in the absence of
additional deformations the total free energy of the gel per chain is:%
\begin{equation}
A^{\text{ch}}=A_{\text{os}}^{\text{ch}}+A_{\text{el}}^{\text{ch}}\simeq
\ k_{B}T\left[  N{\left(  \rho_{\text{eq}}b^{3}\right)  }^{5/4}+{\left(
\rho_{0}/\rho_{\text{eq}}\right)  }^{5/12}\right]  \label{SI_Aeq}%
\end{equation}
and it is minimized at the density\cite{SI_Pa-JLett87}%
\begin{equation}
\rho_{\text{eq}}\simeq\frac{{\left(  \rho_{0}b^{3}\right)  }^{1/4}}%
{b^{3}N^{3/5}}\label{SI_req}%
\end{equation}
corresponding to maximum swelling ratio%
\begin{equation}
\lambda_{\text{eq}}=\left(  \rho_{0}/\rho_{\text{eq}}\right)  ^{1/3}%
\simeq\left(  \rho_{0}b^{3}\right)  ^{1/4}N^{1/5}\label{SI_leq}%
\end{equation}

Note that similar expression for $\lambda_{eq}$ is obtained in mean field
model of a gel with second virial coefficient $B\simeq b^{3}$, see main text.
This conclusion can also be extended to our solution of the image storing
problem.\textbf{\ }Since both elastic (Eq.~(\ref{SI_el1})) and osmotic
(Eq.~(\ref{SI_osm})) terms in the gel free energy are multiplied by the same
scaling factor $\left(  \rho/\rho_{0}\right)  ^{1/4}$ such scaling
renormalization does not change the results obtained for the mean field model.

\section*{SI 2. How does the pattern change in different solvent conditions?}

\subsection*{$\theta$-solvent}

In a $\theta$-solvent the second virial coefficient vanishes ($B=0$) and
equation
\begin{equation}
\frac{\tilde{\rho}\left(  \mathbf{x}\right)  }{\bar{\rho}}=\sum_{i}\lambda
_{i}^{2}\frac{\partial^{2}}{\partial x_{i}^{2}}\int g\left[  \gamma^{-1}%
\cdot\left(  \mathbf{x}-\mathbf{y}\right)  \right]  \tilde{G}\left(
\lambda^{-1}\mathbf{\cdot x}\right)  \frac{d\mathbf{y}}{\prod_{j}\gamma_{j}}
\label{SI_rG}%
\end{equation}
reproduces without distortion affinely stretched initial pattern
\begin{equation}
\tilde{\rho}\left(  \mathbf{x}\right)  /\bar{\rho}=\tilde{G}\left(
\lambda^{-1}\mathbf{\cdot x}\right)  /\bar{G},
\end{equation}
even for anisotropically stretched gels (small deviations from affinity are
expected because of the non-vanishing third virial coefficient)

\subsection*{Poor solvent}

Strong enhancement of monomer density contrast can be obtained by placing the
gel (with fixed boundaries -- otherwise it would collapse) in a poor solvent
with negative second virial coefficient $B<0$. In case of very poor solvent
with
\begin{equation}
\gamma^{2}=\bar{G}\lambda^{2}+k_{B}TB\bar{\rho}^{2}<0 \label{SI_gamma}%
\end{equation}
the gel becomes unstable with respect to formation of domains with different
monomer density\cite{SI_orit}. Below we consider the case of poor solvent close
to $\theta$-conditions with small $B<0$ and positive $\gamma^{2}>0$.

\begin{figure}[tbh]%
\centering
\includegraphics[
height=3.1156in,
width=2.8206in
]%
{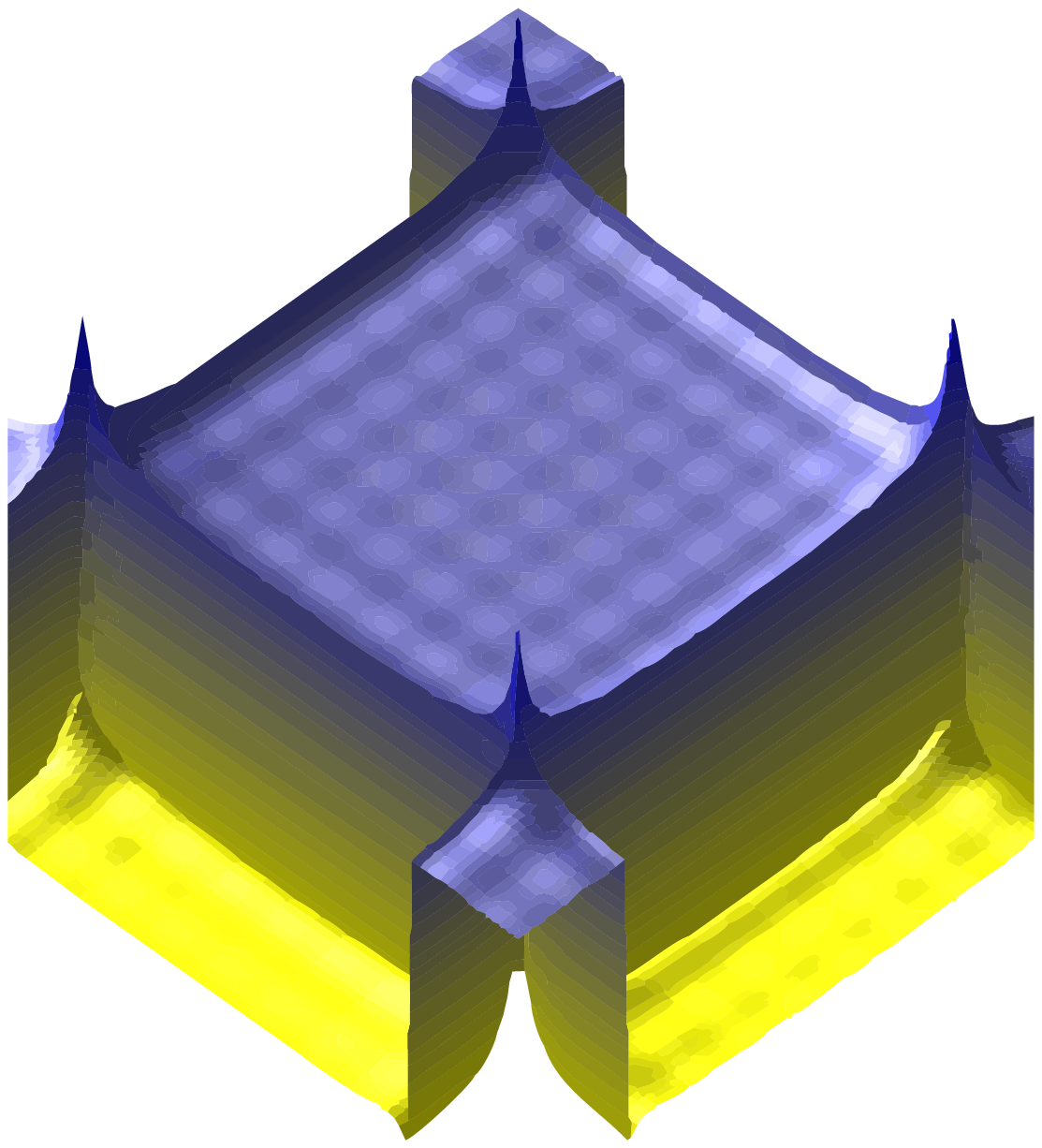}%
\caption{Density profile in poor solvent for initial cross-link concentration
shown in Fig.~1a).}%
\label{SI_poor}%
\end{figure}

At small $\gamma^{2}$ the amplitude of density variations $\tilde{\rho}\left(
\mathbf{x}\right)  $ can be significantly increased because of the small
denominator in equation%
\begin{equation}
\frac{\tilde{\rho}\left(  \mathbf{x}\right)  }{\bar{\rho}}=\frac{\lambda^{2}%
}{\gamma^{2}}\tilde{G}\left(  \frac{\mathbf{x}}{\lambda}\right)  \label{SI_dr}%
\end{equation}
and we have to take into account corrections due to second order in
$\mathbf{u}$ term
\begin{equation}
\Delta\left(  \mathbf{u}\right)  \simeq\sum_{ij}\left(  \frac{\partial u_{i}%
}{\partial x_{i}}\frac{\partial u_{j}}{\partial x_{j}}-\frac{\partial u_{i}%
}{\partial x_{j}}\frac{\partial u_{j}}{\partial x_{i}}\right)  \label{SI_Delta}%
\end{equation}
in expression for monomer density,%
\begin{equation}
\rho\left(  \mathbf{x}\right)  =\frac{\bar{\rho}}{\det\left(  F_{ij}%
^{u}\right)  }, \label{SI_F}%
\end{equation}
where
\begin{equation}
\det\left(  F_{ij}^{u}\right)  =1+\sum_{i}\frac{\partial u_{i}}{\partial
x_{i}}+\Delta\left(  \mathbf{u}\right)  \label{SI_detF}%
\end{equation}
To first order in $\Delta$ we find%
\begin{equation}
\frac{\tilde{\rho}\left(  \mathbf{x}\right)  }{\bar{\rho}}\simeq\frac
{\lambda^{2}}{\gamma^{2}}\left[  \tilde{G}\left(  \frac{\mathbf{x}}{\lambda
}\right)  -\bar{G}\Delta\left(  \mathbf{u}\right)  \right]  \label{SI_corr}%
\end{equation}
where the equilibrium displacement in $\Delta\left(  \mathbf{u}\right)  $ is
determined as%
\begin{equation}
u_{i}\left(  \mathbf{x}\right)  \simeq-\frac{\lambda^{2}}{\gamma^{2}}%
\frac{\partial}{\partial x_{i}}\int g\left(  \mathbf{\mathbf{x}-y}\right)
\tilde{G}\left(  \frac{\mathbf{y}}{\lambda}\right)  d\mathbf{y}%
\end{equation}
We conclude that the correction term in Eq.~(\ref{SI_corr}) enhances the contrast
between the high and the low monomer density regions of the profile
(Fig.~\ref{SI_poor}).
This effect is the mostly pronounced near the corners of the pattern where
several edges converge and it leads to distortion of the otherwise affinely
stretched profile at these points.

\section*{SI 3. Random heterogeneities}

Frozen-in random heterogeneities of network structure can change the image
beyond recognition \cite{SI_orit1}. The free energy of a gel with frozen-in
heterogeities was derived in Ref. \cite{SI_PaRu-Ma96}. The only source of
heterogeneities in the melt with fixed monomer density is statistical
distribution of cross-links in the state of preparation that arises as the
consequence of the random process of cross-linking. This frozen-in
distribution is described by an additional contribution to the free energy:
\begin{equation}
\Delta A_{n}=-\int\sum_{ij}f_{ij}\left(  \mathbf{x}\right)  \frac{\partial
u_{i}}{\partial x_{j}}d\mathbf{x}%
\end{equation}
where $f_{ij}\left(  \mathbf{x}\right)  $ is random Gaussian function of
coordinate $\mathbf{x}$, characterized by correlation function%
\begin{equation}
\overline{f_{ij}\left(  \mathbf{x}\right)  f_{kl}\left(  \mathbf{x}^{\prime
}\right)  }\simeq\left(  k_{B}T\right)  ^{2}\bar{G}\delta\left(
\mathbf{x}-\mathbf{x}^{\prime}\right)  \delta_{ik}\delta_{jl}%
\end{equation}
Comparing the amplitude of frozen-in fluctuations on a scale $R$ with
variations of elastic modulus $\tilde{G}$ on this scale we conclude that the
contribution of frozen-in heterogeneities can be neglected if%
\begin{equation}
\tilde{G}/\bar{G}\gg1/\left(  \bar{c}R^{3}\right)  ^{1/2}%
\end{equation}
where $\bar{c}$ is average cross-link concentration and thus, frozen-in
heterogeities have no influence on large-scale patterns. The suppression of
frozen-in heterogeneities of monomer density is due to strong overlap of
network chains \cite{SI_RuCo-PolPH03}.

\end{document}